\documentclass[twocolumn, aps,pra, superscriptaddress,longbibliography]{revtex4-2}

\usepackage{bm}
\usepackage{appendix}
\usepackage{graphicx}
\usepackage{amsmath}
\usepackage{amssymb}
\usepackage{color}
\usepackage{lipsum}
\usepackage{ulem}
\usepackage[colorlinks=true,citecolor=blue,urlcolor=blue]{hyperref}

\begin{document}
\title{Dynamic generation of superflow in a fermionic ring through phase imprinting}
\date{\today}

\author{Ke-Ji Chen}
\email{chenkeji2010@gmail.com}
\affiliation{Key Laboratory of Quantum States of Matter and Optical Field Manipulation of Zhejiang Province, Department of Physics, Zhejiang Sci-Tech University, 310018 Hangzhou, China}

\author{Wei Yi}
\affiliation{CAS Key Laboratory of Quantum Information, University of Science and Technology of China, Hefei 230026, China}
\affiliation{CAS Center For Excellence in Quantum Information and Quantum Physics, Hefei 230026, China}
\affiliation{Hefei National Laboratory, University of Science and Technology of China, Hefei 230088, China}

\author{Fan Wu}
\email{t21060@fzu.edu.cn}
\affiliation{Fujian Key Laboratory of Quantum Information and Quantum Optics, College of Physics and Information Engineering, Fuzhou University, Fuzhou, Fujian 350108, China}

\begin{abstract}
We study the dynamic generation of persistent current by phase imprinting fermionic atoms in a ring geometry at zero temperature. Mediated by the pairing interaction, the Fermi condensate dynamically acquires a quantized current by developing azimuthal phase slips, as well as density and pairing-order-parameter depletions. Resorting to the Bogolioubov-de Gennes formalism, we investigate the time evolution of the transferred total angular momentum and the quantized superfluid current throughout the phase-imprinting process.
This enables a detailed self-consistent analysis of the impact of interaction, as well as different initial pairing states, on the superflow formation,  in contrast to previous theoretical analysis based on the Gross-Pitaevskii equation with artificially imposed phases. In particular, we show that, as the interaction strength increases,
the azimuthal density distribution becomes less susceptible to the phase imprinting potential, leading to a smaller quantized current under the same imprinting parameters. Our results offer  microscopic insights into the dynamic development of superflow in the phase-imprinting process, and are helpful for the ongoing experimental effort.
\end{abstract}

\maketitle
\section{introduction}
Superflow (or persistent current), in ring geometries threaded by a magnetic field, is central to the identification and application of the macroscopic quantum coherence in superconductors~\cite{Tinkham-04, Annett-06, Akkermans-07, Tsai-99, Wilhelm-08, Fairbank-61, Doll-61}. The
long-lived current, following the quantization of the magnetic flux through the ring, is also quantized, dictated by the phase winding of the pairing wave function under the vector potential along the perimeter of the ring~\cite{Yang-61,Bloch-65, Fairbank-61, Doll-61,Onsager-61}.
In charge-neutral cold atoms, persistent currents can also be induced, in either Bose-Einstein or Fermi condensates, by imposing synthetic gauge fields~\cite{Lin-09, Dalibard-11,Spielman-14}. This can be achieved, for instance, through rotation~\cite{Cornell-99, Madison-00, Ketterle-01, Fetter-09, Cooper-08, Ketterle-05, Piazza-09, Phillips-13, Cornell-14, Wright-22},
or by enforcing laser-assisted gauge potentials~\cite{Phillips-06, Phillips-07, Phillips-11, Hadzibabic-13, Lin-18,Jiang-19, Pu-15,Sun-15,Qu-15,Chen-16,Hu-19,Chen-19,Pu-20,Duan-20,Chen-20,Wang-21,Chen-22, Han-22, Peng-22}.
These practices open up intriguing avenues for studying the generation and dissipation of superflow in the highly controllable environment of neutral atoms.
Compared to the light-assisted synthetic gauge fields, the recently demonstrated
phase-imprinting techniques offer a more straightforward route toward persistent current in cold atoms~\cite{Hadzibabic-12, Kumar-18, Roati-22, Roati24}.
For instance, superflow of Bose-condensed atoms can be excited by subjecting the condensate to
light shift with an azimuthal gradient~\cite{Hadzibabic-12, Kumar-18}.
In a similar spirit, phase winding of the Fermi superfluid is observed when fermionic atoms in a ring trap are subject to a light-assisted phase gradient~\cite{Roati-22}.
Here the dynamic  generation of  superflow in fermions is particularly intriguing: since phase imprinting is a single-particle process, the dynamic transfer of angular momentum from the light beams to the Cooper pairs necessarily involves pairing interaction, whose role in the process is yet to be clarified.

\begin{figure}[t]
\begin{center}
\includegraphics[width=0.46\textwidth]{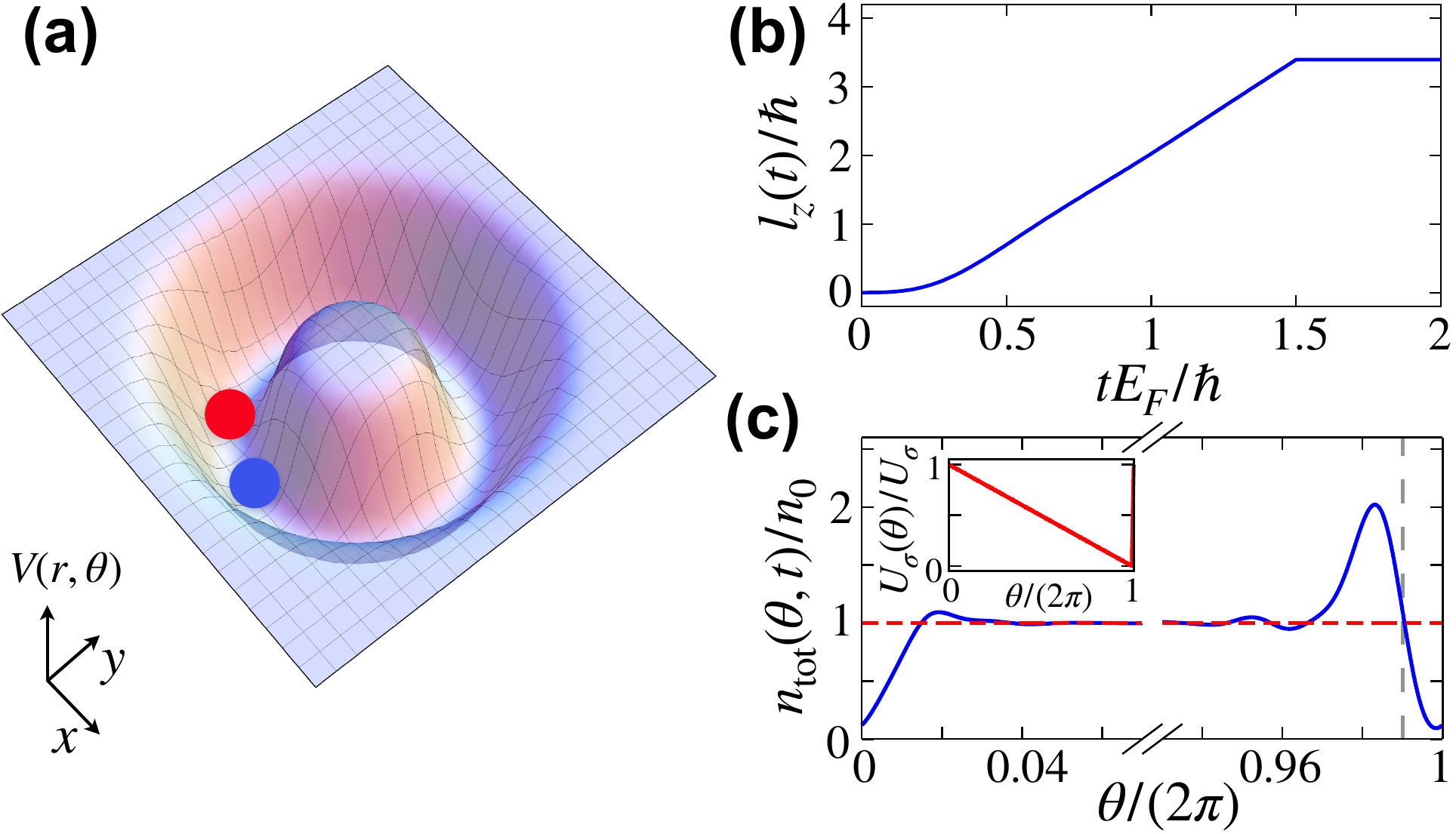}
\caption{(a) Schematic illustration of $V(r,\theta)$, which is responsible for the ring trap.  The red and blue solid dots denote atoms of different spin species. (b) Evolution of the angular momentum $l^{\rm tot}_z(t)/\hbar$ (blue solid curve)  of a noninteracting Fermi gas. (c) Angular density distribution $n_{\rm tot}(\theta, t)/n_0$,  with $n_{\rm tot}(\theta,t)=\sum_{\sigma}n_{\sigma}(\theta,t)$, for $t E_F/\hbar=0$ (red dashed curve) and $t E_F/\hbar=0.5$ (blue solid curve), respectively. The inset shows the profile of  $U_{\sigma}(\theta)/U_{\sigma}$ (red solid curve), the vertical dashed line is located at $\theta/(2\pi)=(2\pi-\Delta \theta)/(2\pi)$.  In (b) and (c),  the parameters are $U_{\uparrow}=U_{\downarrow}=10E_F$, $\Delta \theta=0.01 \pi$, $\tau E_F/\hbar=1.5$, $h=0$, $k_F R=15$ with $k_F$ the Fermi vector, and $n_0=N/(2\pi)$.}
\label{Fig1}
\end{center}
\end{figure}

\begin{figure}[t]
\begin{center}
\includegraphics[width=0.48\textwidth]{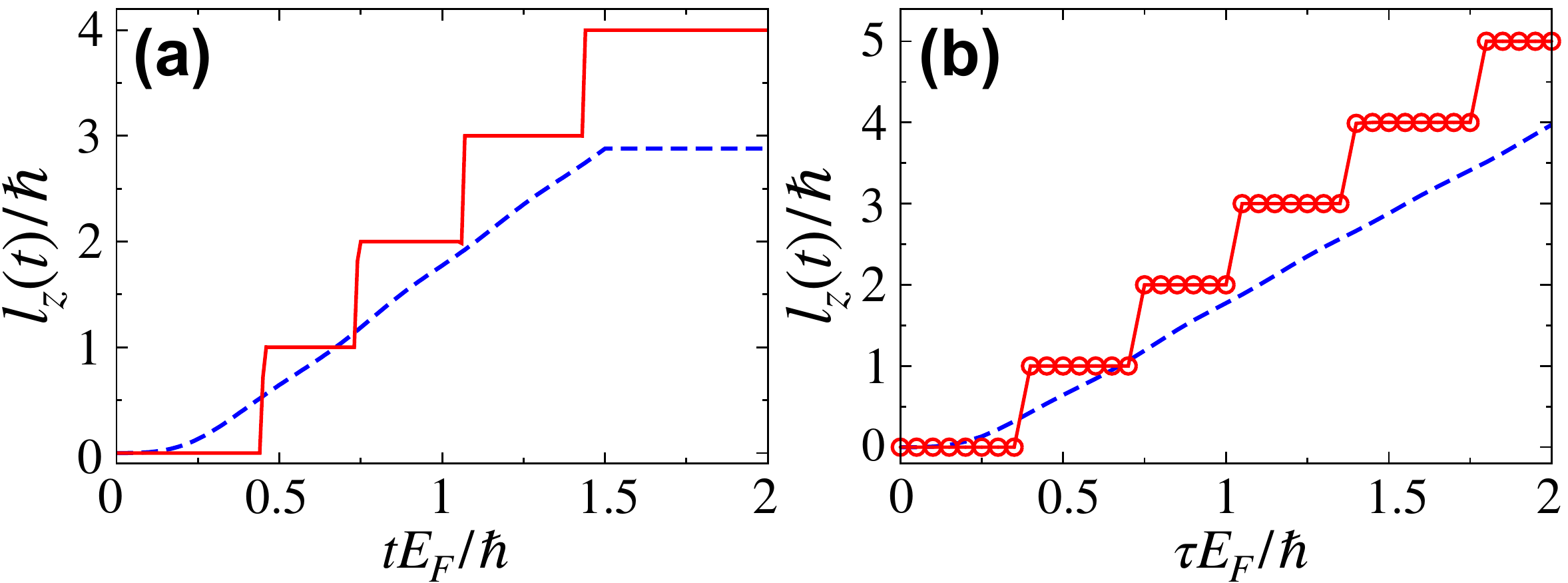}
\caption{(a) Evolution of  $l^{\rm tot}_z(t)/\hbar$ (blue dashed curve) and $l^{\Delta}_{z}(t)/\hbar$ (red solid curve) with a fixed phase-imprinting time $\tau E_F/\hbar=1.5$. (b) The injected angular momenta $l^{\rm tot}_z(t)/\hbar$ (blue dashed curve) and $l^{\Delta}_{z}(t)/\hbar$ (red solid curve with circles) under different $\tau$ at long times. Here, $E_B/E_F=-5$;  other parameters are the same as those in Fig.\ref{Fig1}.}
\label{Fig2}
\end{center}
\end{figure}

\begin{figure*}[t]
\begin{center}
\includegraphics[width=0.88\textwidth]{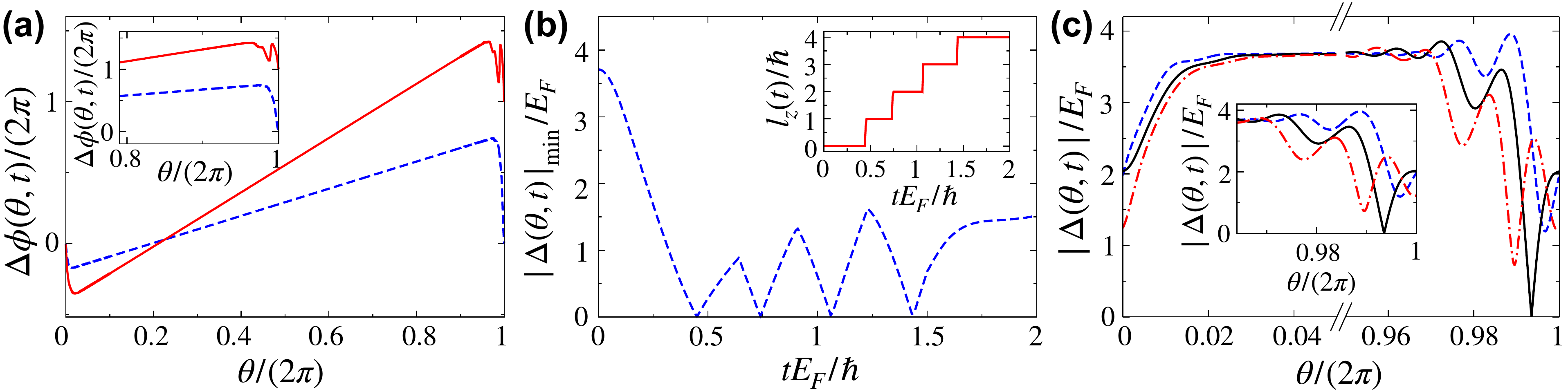}
\caption{(a) Evolution of $\Delta \phi(\theta,t)/(2\pi)$ for $tE_F/\hbar=0.3$ (blue dashed curve) and $t E_F/\hbar=0.6$ (red solid curve) with a fixed $\tau E_F/\hbar$. The inset shows a zoom-in view of $\Delta \phi(\theta,t)/(2 \pi) $ for $\theta/(2\pi) \in [0.8,1]$.  (b) Evolution of $|\Delta(\theta,t)|_{\rm{ min}}/E_F$ (blue dashed curve) with a fixed $\tau E_F/\hbar$. The inset shows the evolution of  $l^{\Delta}_z(t)/\hbar$ (red solid curve). (c) Profiles of $|\Delta(\theta,t)|/E_F$ at $tE_F/\hbar=0.3$ (blue dashed curve), $t E_F/\hbar=0.45$ (black solid curve) and $tE_F/\hbar=0.6$ (red dash-dotted curve) with $\tau E_F\hbar=1.5$.  The inset shows a zoom-in  view of the profiles of $|\Delta(\theta,t)|/E_F$ as $\theta$ approaches $2\pi$. Other parameters are the same as those in Fig.\ref{Fig2}.}
\label{Fig3}
\end{center}
\end{figure*}

In this work, we study the dynamic generation of superflow in a ring-shaped Fermi gas under phase imprinting. The dynamic process was theoretically analyzed based on
the Gross-Pitaevskii equation~\cite{Roati-22, Roati-23},  which should only apply in the Bose-Einstein-condensate (BEC) regime of the Fermi condensate. To provide a more general description as well as microscopic insights on the dynamic superflow generation,
we adopt a Bogoliubov-de Gennes (BdG) formalism,  where both the phase imprinting process and the superflow generation naturally arise in a self-consistent manner. We focus on the transfer of angular momentum and the emergence of quantized current (Fig.\ref{Fig2}) throughout the imprinting process, revealing rich dynamic features, including the density depletion, order-parameter phase slip, and the dynamical phase transitions.  We demonstrate that, first (Fig.\ref{Fig1}), the density depletion induced by the phase imprinting potential leads to the increase in the transferred total angular momentum. Second (Fig.\ref{Fig3}), consistent with the superfluid nature of the pairing state, the phase winding of the pairing order parameter emerges through phase slips and order-parameter depletions in the azimuthal direction, driving a dynamical phase transition from a 
zero  to a  nonzero circulation state through the phase imprinting technique.
More importantly (Fig.\ref{Fig4}), we find that both the total angular-momentum transfer and the quantized-current generation are hindered under stronger interactions. This is because the system becomes less susceptible to density modulations under stronger interactions, whereas density depletions are an inevitable concomitant of those in the order parameter.
On the other hand, when the Fermi gas is initialized in an angular Larkin-Ovchinnikov (LO) state~\cite{LO-65}, where the pairing order parameter has an azimuthal amplitude modulation, the quantized current generation is also suppressed. This is because the amplitude modulation of the order parameter accommodates part of the transferred angular momentum, leaving less for the quantized phase winding.
Our results provide microscopic details for the dynamic phase imprinting in Fermi superfluids, and are helpful for devising  more efficient imprinting protocols.

%%%%%%%%%%%%%%%%%%%%%%%%%%%%%%%%%
%%%%%%%%%%%%%%%%%%%%%%%%%%%%%%%%

\section{model}
As illustrated in Fig.\ref{Fig1}(a), we consider a two-component Fermi gas with atom mass $M$ confined by an annular potential~\cite{Roati-22}
  \begin{eqnarray}
V(r,\theta)  =  \sum_{j=1,2}V_0 \left (\tanh \left[\frac{(-1)^{j}(r-R_{j})}{d}\right]+1\right)
\label{V(r,theta)}
\end{eqnarray}
in the $x-y$ plane, and by a potential $V(z)=M \omega^2_z z^2/2 $ in the $z$ direction.
Here $V_0$, $R_{1}(R_2)$ denote the trapping strength, inner (outer) radius of  $V(r,\theta)$, respectively, $d$ is a parameter, and $\omega_z$ is the trapping frequency along the $z$ axis.
We consider the experimentally relevant case~\cite{Roati-22} with $\hbar \omega_z$ being the largest energy scale, $d\ll R_{j}$, and $R_{1} \approx R_{2}$. Under these restrictions, atomic motion in the axial and radial directions is suppressed, resulting in a ring-shaped Fermi gas with a radius $R$, where $R \equiv (R_1+R_2)/2$.

Phase imprinting is realized through an angular potential $U_{\sigma}(\theta)$, with
\begin{eqnarray}
U_{\sigma}(\theta)=\left\{\begin{array}{l}  U_{\sigma} \Big [1-\frac{\theta}{2\pi -\Delta \theta}\Big] , \,\,  \theta \in [0,2\pi- \Delta \theta ],\\
\\
\frac{U_{\sigma}}{\Delta \theta}\Big[\theta -(2\pi-\Delta \theta)\Big], \,\, \theta \in (2 \pi-\Delta \theta, 2\pi).  \end{array}\right.
\label{U_sigma}
\end{eqnarray}
Here $U_{\sigma}$ ($\sigma=\uparrow, \downarrow$) is the spin-dependent potential depth, and $\Delta \theta \ll 2\pi$. Since $[U_{\sigma}(\theta), \hat{L}_z] \neq 0$ (here $\hat{L}_z \equiv -i \hbar{\partial }/{\partial \theta}$), the angular potential $U_{\sigma}(\theta)$ plays a significant role in introducing angular momentum to the Fermi gas.
Specifically, the phase imprinting process is realized by turning on $U_{\sigma}(\theta)$ at $t=0$ for a duration of $\tau$.

We start by characterizing the phase imprinting process in a noninteracting %two-component
Fermi gas, which provides a useful context for that in a Fermi condensate.
The time-dependent Hamiltonian of the system can be expressed as
$ H_0(t) =  \sum_{\sigma} \int d\theta \psi^{\dag}_{\sigma}(\theta,t){ \cal H}_{\sigma}(\theta,t)\psi_{\sigma}(\theta,t) $,
with $\psi_{\sigma}(\theta,t)$ the fermion field operator for the spin species $\sigma$, and
\begin{align}
{\cal H}_{\sigma}(\theta, t)={\cal H}_{\sigma}(\theta)+U_{\sigma}(\theta) \vartheta (\tau-t).
\label{K_sigma}
\end{align}
Here ${\cal H}_{\sigma}(\theta) =  -\hbar^2/(2MR^2) \partial^2 /\partial \theta^2 -\mu_{\sigma}$, where $\mu_{\sigma}$, the spin-dependent chemical potentials,  are parameterized by $\mu$ and $h$ through $\mu_{\sigma}=\mu+s h$, with $s=+1(-1)$ for $\sigma=\uparrow(\downarrow)$. $\vartheta(x)$ is  the Heaviside step function. For now, we focus on an unpolarized Fermi gas with $h=0$. We assume that the Fermi gas is initially in the ground state of the Hamiltonian $H_0= \sum_{\sigma} \int d\theta \psi^{\dag}_{\sigma}(\theta) {\cal H}_{\sigma}(\theta) \psi_{\sigma}(\theta)$, and calculate the time evolution of $l^{\rm tot}_z(t)$, where $l^{\rm tot}_z(t)=L^{\rm tot}_z(t)/ N_{\rm p}$, and $L^{\rm tot}_{z}(t)=\sum_{\sigma}L^{\sigma}_z(t)$ is the total angular momentum, with $L^{\sigma}_{z}(t) \equiv \langle \psi_{\sigma}(\theta,t) |\hat{L}_z |\psi_{\sigma}(\theta,t)\rangle$ and $N_{\rm p}=N/2$.

As illustrated in Fig.\ref{Fig1}(b), when $t \leqslant \tau$, $l^{\rm tot}_z(t)$ increases from zero and remains conserved for $t > \tau$. Such a behavior can be understood from the equation of motion for $L^{\sigma}_z(t)$
\begin{align}
\frac{d}{dt}L^{\sigma}_z(t)=U_{\sigma} \Big({\bar n}^{L}_{\sigma}(t)-\bar{n}^{R}_{\sigma}(t)\Big)\vartheta(\tau-t),
\label{dLz_time}
\end{align}
where $\bar{n}^{L}_{\sigma}(t)$ and $\bar{n}^{R}_{\sigma}(t)$ represent the average densities  of the corresponding spin component for $\theta \in [0, 2\pi-\Delta \theta]$ and $\theta \in (2\pi-\Delta \theta,2\pi)$, respectively. Specifically, we have
\begin{align}
\bar{n}^{L}_{\sigma}(t) &=  \frac{1}{2\pi-\Delta \theta} \int ^{2\pi-\Delta \theta}_0  d \theta  n_{\sigma}(\theta,t), \label{nL}\\
\bar{n}^{R}_{\sigma}(t)  &=  \frac{1}{\Delta \theta} \int^{2\pi}_{2\pi-\Delta \theta} d \theta n_{\sigma}(\theta,t),
\label{nR}
\end{align}
where $n_{\sigma}(\theta,t)=\psi^{\dag}_{\sigma}(\theta,t) \psi_{\sigma}(\theta,t)$ [see Appendix \ref{Motion} for details].
As shown in Fig.\ref{Fig1}(c), at early times of the evolution, the density distribution develops a depletion near $\theta\approx 2\pi$, where the phase-imprinting  potential $U_\sigma(\theta)$ changes rapidly [see the inset of Fig.\ref{Fig1}(c)].
Such a depletion leads to an uneven density distribution in the azimuthal direction, with
$\bar{n}^{L}_{\sigma}(t) > \bar{n}^{R}_{\sigma}(t)$.
According to Eq.(\ref{dLz_time}), for $t \leqslant \tau$,
this disparity results in an increase in the system's total angular momentum.
Hence, the phase imprinting alters the total angular momentum of a noninteracting Fermi gas by developing uneven density distributions in the azimuthal direction. The transferred angular momentum is not quantized in general.

The picture above is qualitatively modified in the presence of pairing interactions.
We consider an $s$-wave interaction between the two spin species, so that the
full Hamiltonian reads
$H(t)=H_0(t)+ H_{\rm int}(t)$, where $ H_{\rm int}(t)=-g\int d\theta \psi^{\dag}_{\uparrow}(\theta,t)\psi^{\dag}_{\downarrow}(\theta,t)\psi_{\downarrow}(\theta,t)\psi_{\uparrow}(\theta,t)$,
and $g$ is the bare interaction strength, renormalizable through the two-body binding energy $E_B$ in one dimension [see Appendix \ref{renormal} for details].
To provide a microscopic insight into the superflow generation through phase imprinting, we employ the BdG formalism, concentrating on the dynamics of the angular momentum transfer and the development of quantized currents. This approach is epitomized by the time-dependent BdG equations
\begin{align}
i \hbar \frac{\partial }{\partial t} \left[\begin{array}{c}u_{\uparrow n}(\theta,t) \\ v_{\downarrow n}(\theta,t)\end{array}\right]
=\left[\begin{array}{cc} {\cal H}_{\uparrow}(\theta,t) & \Delta(\theta,t) \\  \Delta^{\ast}(\theta,t) & -{\cal H}^{\ast}_{\downarrow}(\theta,t)\end{array}\right]
\left[\begin{array}{c}u_{\uparrow n}(\theta,t) \\ v_{\downarrow n}(\theta,t)\end{array}\right],
\label{td-BdG}
\end{align}
where $u_{\sigma n}(\theta,t)$ and $v_{\sigma n}(\theta, t)$ are the Bogoliubov coefficients, and the time-dependent pairing order parameter
$ \Delta(\theta,t) =g\langle \psi_{\uparrow}(\theta,t)\psi_{\downarrow}(\theta,t)\rangle $.
It follows that, for an initial state with a fixed total particle number $N$,
the time evolution of
$\Delta(\theta,t)$ and $l^{\rm tot}_z(t)$ are determined self-consistently
from Eq.(\ref{td-BdG}) [see Appendix \ref{TBdG} for details].

%%%%%%%%%%%%%%%%%%%%%%%%%%%%%%%%%%%
%%%%%%%%%%%%%%%%%%%%%%%%%%%%%%%%%%

%%%%%%%%%%%%%%%%%%%%%%%%%%%%%%%%%%%
%%%%%%%%%%%%%%%%%%%%%%%%%%%%%%%%%%

%%%%%%%%%%%%%%%%%%%%%%%%%%%%%%%%%%
\section{superflow generation and dynamic transition}
With pairing interactions, the impact of the phase imprinting generally depends
on the system parameters such as the interaction strengths, the spin-dependent potentials $U_{\sigma}$, and the initial states.

We first study the emergence of a quantized current in the pairing-order parameter by considering the simplest scenario:
the system is initialized in the ground Bardeen-Cooper-Schrieffer (BCS) pairing state with $h=0$, and $U_{\uparrow}=U_{\downarrow}$. Figure~\ref{Fig2}(a) shows the dynamics of the total transferred angular momentum $l^{\rm tot}_z(t)$ per fermion pair, which continually increases from zero and is not quantized. This is understandable, since $l^{\rm tot}_z(t)$ contains contributions from both the phase and amplitude modulation of the pairing wave function.

To further elucidate the quantized current component, we focus on the pairing order parameters of the system $\Delta(\theta,t)$, which is expressed as
$\Delta(\theta,t)  =  |\Delta(\theta,t)|e^{i \phi(\theta,t)}$ and satisfies $\Delta(0, t)=\Delta(2\pi,t)$.
This gives rise to $\phi(2\pi,t)-\phi(0, t)=2\pi \kappa$, where $\kappa$ is the winding of the phase of $\Delta(\theta,t)$: $\kappa =0 $ and $\kappa \neq 0$ correspond to states without and with 
quantized current, respectively. 
In the following, we label the state without quantized current as the BCS state.
We thus define the angular momentum associated with the phase of $\Delta(\theta,t)$
\begin{eqnarray}
l^{\Delta}_{z}(t)= \frac{\hbar}{2\pi} \int^{2\pi}_0 d\theta \frac{\partial \phi(\theta,t)}{\partial \theta},
\label{lz-Delta}
\end{eqnarray}
which is the quantized component of the current. As illustrated in
Fig.\ref{Fig2}(a), three key features of $l^{\Delta}_z(t)$ are identified. First, $l^{\Delta}_z(t)$ is quantized as expected, and provides a useful indicator for the superflow generation.
Second, $l^{\Delta}_z(t)$ can jump between different quantized values $\kappa \hbar$, during the imprinting process with $t\leq \tau$.  Third, $l^{\Delta}_z(t)$ stabilizes for $t > \tau$, indicating the robustness of a  persistent
current state.  Unlike in Ref.~\cite{Roati-22, Roati-23}, where the quantized current  can decay over time
due to vortex emission, our one-dimensional system constrains the radial and azimuthal degrees of freedom, preventing vortex formation. As a result, the quantized current in our system exhibits remarkable stability over time after the phase imprinting. Figure \ref{Fig2}(b) shows the dependence of the final stable $l^{\rm tot}_z(t)$ and $l^{\Delta}_z(t)$ on the imprinting time $\tau$. Based on Fig.\ref{Fig2}(b), 
quantized current states with specific winding numbers can be prepared by tuning the imprinting time $\tau$.

The abrupt jumps in the evolution of $l^{\Delta}_z(t)$ correspond to dynamic transitions between the BCS state and different quantized current states.
To further understand these jumps, we calculate the phase evolution of $\Delta(\theta,t)$.
In Fig.\ref{Fig3}(a), we show the numerically evaluated
$\Delta \phi(\theta,t)  =  \int^{\theta}_0 d\theta  \partial \phi(\theta,t) /\partial \theta $ at different times of the phase imprinting [see Appendix \ref{phase of Delta} for details]. At $tE_F/\hbar=0.3$, we have $\Delta \phi(2\pi,t)=0$ with $l^{\Delta}_z(t)/\hbar=0$. By contrast, when $t E_F/\hbar =0.6$, we have $\Delta \phi(2\pi,t)=2\pi$ with $l^{\Delta}_z(t)/\hbar=1$.
Therefore, in between the two time points, a dynamic transition between a BCS state with $\kappa=0$ and a  quantized current state with $\kappa=1$ necessarily occurs through a phase slip.
In Fig.\ref{Fig3}(b), we show the time evolution of the minimum order parameter in the angular direction (labeled as $|\Delta(\theta,t)|_{\rm min}$ and shown in blue dashed curve), as well as the evolution of $l^{\Delta}_z(t)$ (red solid curve in the inset). The jumps in the quantized angular momentum occur at  locations where $|\Delta(\theta,t)|_{\rm min}=0$.
Further, in Fig.\ref{Fig3}(c), we confirm the results above by showing the profile of $|\Delta(\theta,t)|$ along $\theta$ at different times. Importantly,
when $ tE_F/\hbar \approx 0.45$, a nodal point emerges in $|\Delta(\theta,t)|$,  with $|\Delta(\theta,t)|$ completely depleted when $\theta\approx 2\pi$, which gives rise to the abrupt jump in the winding number.
Thus, the emergence of the nodal point in $|\Delta(\theta,t)|$ serves as an indicator for the dynamic transition.  From the detailed analysis above, we conclude that the superflow generation  arises from the phase slip, accompanied by the depletion of the order parameter, signifying a dynamical phase transition from a BCS state to a   quantized current state through the phase imprinting process.

%%%%%%%%%%%%%%%%%%%%%%%%%%%%%%%%%%
\begin{figure}[t]
\begin{center}
\includegraphics[width=0.48\textwidth]{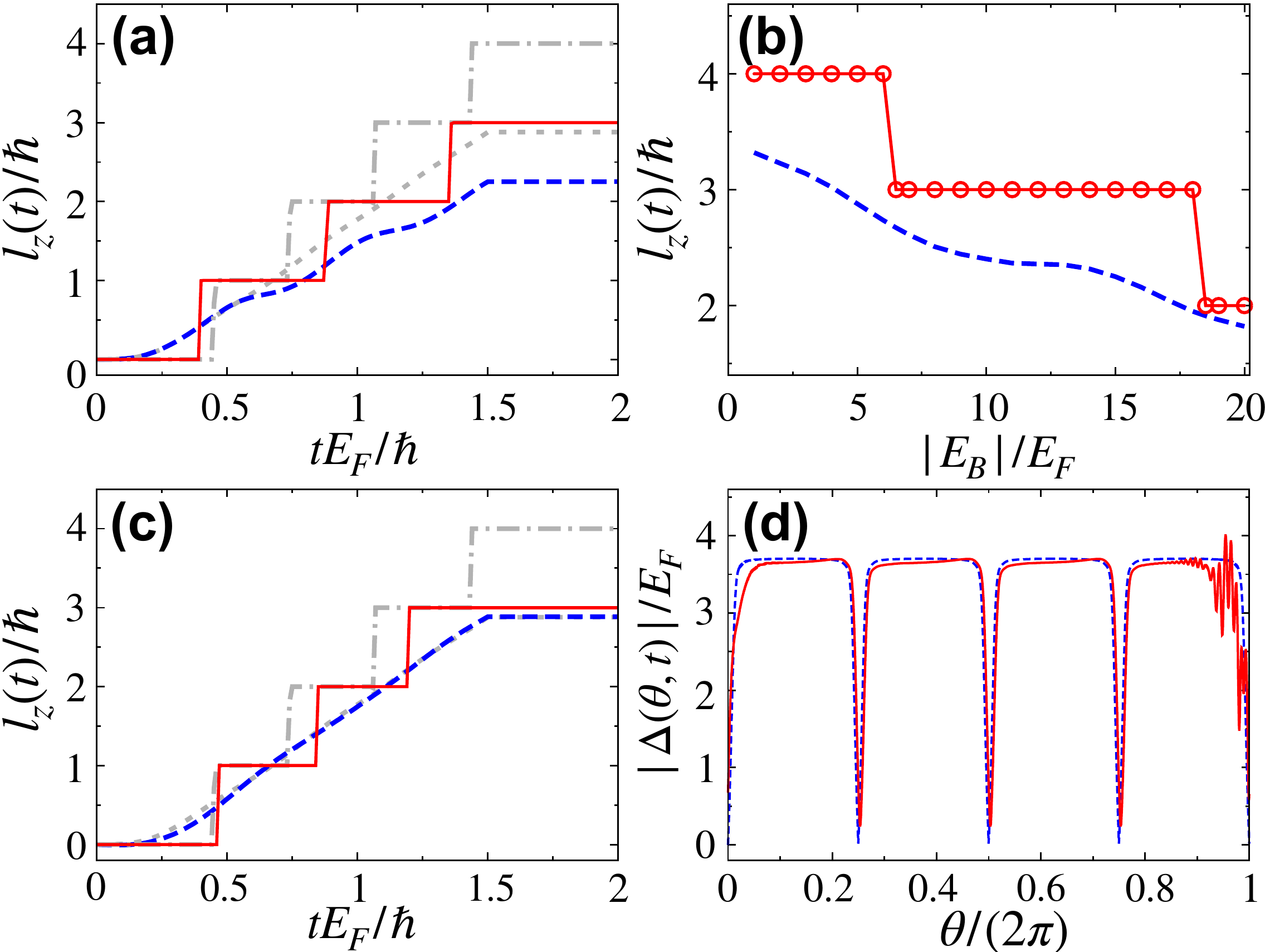}
\caption{(a) Evolution of $l^{\rm tot}_z(t)/\hbar$ and $l^{\Delta}_z(t)/\hbar$  under different interaction strengths with a fixed $\tau E_F\hbar$. The blue dashed (gray dotted) and red solid (gray dash-dotted) curves denote $l^{\rm tot}_z(t)/\hbar$ and $ l^{\Delta}_z(t)/\hbar$, respectively, for  $E_B/E_F=-15$ ($E_B/E_F=-5$).
(b) Relation between the final $l^{\rm tot}_z(t)/\hbar$ ($l^{\Delta}_z(t)/\hbar)$ and $|E_B|$.  The blue dashed curve and the red solid curve with circles denote the final $l^{\rm tot}(t)/\hbar$ and $l^{\Delta}_z(t)/\hbar$, respectively. (c) Evolution of $l^{\rm tot}_z(t)/\hbar$ and $l^{\Delta}_z(t)$ under different initial states. The blue dashed (gray dotted) and red solid (gray dash-dotted) curves denote $l^{\rm tot}_z(t)/\hbar$ and $l^{\Delta}_z(t)/\hbar$ respectively, for a condensate initialized in the LO (BCS) state.  (d) Evolution of $|\Delta(\theta,t)|/E_F$  when $tE_F/\hbar=0$ (blue dashed curve) and $t E_F/\hbar=1.5$ (red solid curve) with the LO state as the initial state. In (c) and (d), $h/E_F=1.2$.  Other parameters are the same as those in Fig.\ref{Fig2}.}
\label{Fig4}
\end{center}
\end{figure}
%%%%%%%%%%%%%%%%%%%%%%%%%%%%%%%%%%

%%%%%%%%%%%%%%%%%%%%%%%%%%%%%%%%%%
\section{Impact of interaction and initial states}
We now study the impact of interaction strength and different initial states on
the phase imprinting process. Effects induced by the spin-dependent potentials $U_{\sigma}$, such as the inter-species angular-momentum exchange, can be found in Appendix E.

We first choose the BCS state as the initial state, and compare $l^{\rm tot}_z(t)$ [$l^{\Delta}_z(t)$] under different interaction strengths,
characterized by $E_B$ through the renormalization condition.
In Fig.\ref{Fig4}(a), we observe that stronger interactions (larger $|E_B|$) suppress $l^{\rm tot}_z(t)$. Qualitatively, this is because interactions favor a homogenous density distribution along $\theta$. Thus, under stronger interactions, the system acquires a smaller  $\bar{n}^{L}_{\sigma}(t)-\bar{n}^{R}_{\sigma}(t)$, which then suppresses $l^{\rm tot}_z(t)$ [see Appendix \ref{Diffint} for details].
On the other hand, the dynamic consequence of increasing the interaction strength becomes subtle for $l^{\Delta}_z(t)$.
As shown in Fig.\ref{Fig4}(a), at short times, we find that the jump time of $l^{\Delta}_z(t)$ appears earlier for strong interactions, which can be understood as interactions favor a homogeneous $\Delta(\theta,t)$. It follows that, under stronger interactions, the amplitude fluctuation of $\Delta(\theta,t)$ carries less angular momentum, so that a larger proportion of the imprinted angular momentum is distributed to the phase of $\Delta(\theta,t)$. At long evolution times, the total transferred angular momentum under strong interactions decreases compared to weak interactions, leading to a suppressed $l^{\Delta}_z(t)$ as shown in Fig.\ref{Fig4}(a). Figure \ref{Fig4}(b) shows the relation between the final $l^{\rm tot}_z(t)$ and $l^{\Delta}_z(t)$ with respect to $|E_B|$. Consistent with the above analysis, we find smaller $l^{\rm tot}_z(t)$ and $l^{\Delta}_z(t)$ under larger $|E_B|$.

We then calculate  $l^{\rm tot}_z(t)$ and $l^{\Delta}_z(t)$ for a Fermi gas initialized in the Larkin-Ovchinnikov (LO) state.
As illustrated in Fig.\ref{Fig4}(c),
we find that $l^{\rm tot}_z(t)$ under both initial states are similar. By contrast, for $l^{\Delta}_z(t)$, we observe $l^{\Delta}_z(t)$ is suppressed when the system is initialized in the LO state, especially at long times.
This can be explained by the periodic density modulation in the $\theta$ direction of the LO state. As shown in Fig.\ref{Fig4}(d), the amplitude of $\Delta(\theta,t)$ can carry more angular momentum compared to the BCS state, which then suppresses $l^{\Delta}_z(t)$.

\section{Conclusions}
To summarize, we investigate the superflow generation and dynamic transitions induced by angular phase imprinting in a Fermi condensate from a microscopic perspective.
We show that dynamic transitions can be induced between the  BCS
 pairing state and different quantized
current states through the phase imprinting technique. Our microscopic approach reveals that transitions between states with different 
quantized current states
are induced by pairing-order-parameter depletions in the angular direction through phase slips. We further reveal the impact of interaction strength and initial pairing states in the phase imprinting process.
Our results provide microscopic understandings for the recent experimental demonstration of phase imprinting in Fermi condensates~\cite{Roati-22}, and
are the basis for improving the current protocol.
For future studies, it would be interesting to more thoroughly investigate the phase imprinting process in polarized Fermi gases, where the rich phases may play a more significant role.
It would also be intriguing to extend our formalism to Bose-Fermi mixtures, where phase imprinting has yet to be demonstrated experimentally.

{\it {Note added}.}
Recently, a related preprint appeared~\cite{Xhani-24},  where  the complementary problem of supercurrent decay was studied.

%The acknowledgments part.

\section{Acknowledgments}
We acknowledge fruitful discussions with Pen Zou, Zi Cai, Xiaobing Luo,  Dongyang Yu,  and Yajiang Chen. This work is supported by  the Natural Science Foundation of China (Grants No. 12104406, No. 12374479 and No. 12204105) and  the Innovation Program for Quantum Science and Technology (Grant No. 2021ZD0301904).
K.C. acknowledges support from the startup grant of Zhejiang Sci-Tech University (Grant No. 21062338-Y).   F. W. is also supported by  the Natural Science Foundation of Fujian Province (Grant No. 2022J05116).

\section*{APPENDIX}
In these appendices, we provide details on the derivation of the equation of motion for the angular momentum, the renormalization of bare interaction strength $g$, the formalisms of both the dynamical and static BdG equations,   the calculation of the phase of $\Delta(\theta,t)$, and the effects induced by $U_{\sigma}$ and interactions.

\appendix
\section{Motion equations of angular momentum}
\label{Motion}
%\section{Motion equations of angular momentum}
\subsection{Noninteracting case}
The equation of motion for the angular momentum is derived as follows. Before addressing the interacting Fermi system,  we first consider a noninteracting Fermi gas with $N$ particles. We start from the definition of spin-dependent angular momentum $L^{\sigma}_z(t)$, that is, $L^{\sigma}_z(t)=\langle \psi_{\sigma}(\theta,t)| \hat{L}_z |\psi_{\sigma}(\theta,t)\rangle$ with ${\hat L}_z \equiv-i \hbar \partial /\partial \theta$, and the time-dependent Schr\"{o}dinger equation $i \hbar  \partial \psi_{\sigma}(\theta,t)/ \partial t={\cal H}_{\sigma}(\theta,t)\psi_{\sigma}(\theta,t)$,  where ${\cal H}_{\sigma}(\theta,t)={\cal H}_{\sigma}(\theta)+U_{\sigma}(\theta)\vartheta(\tau-t)$ as shown in the main text.  The equation of motion for $L^{\sigma}_z(t)$ can be expressed as
\begin{align}
\label{dLz}
\frac{d}{dt}L^{\sigma}_z(t)= \frac{1}{i \hbar} \langle \psi_{\sigma}(\theta,t) | \Big[{\hat L}_z, {\cal H}_{\sigma}(\theta,t)\Big]|\psi_{\sigma}(\theta,t)\rangle.
\end{align}

Since $[{\hat L}_z,{\cal H}_{\sigma}(\theta)]=0$ and $U_{\sigma}(\theta)$ is as  shown in Eq.(\ref{U_sigma}), we have
\begin{eqnarray}
[\hat {L}_z, U_{\sigma}(\theta)]=\left\{\begin{array}{l}  i \hbar  \Big[\frac{U_{\sigma}}{2\pi -\Delta \theta} \Big] , \,\, \theta \in [0,2\pi- \Delta \theta ],\\
\\
-i \hbar \Big[\frac{U_{\sigma}}{\Delta \theta}\Big], \,\, \theta \in (2 \pi-\Delta \theta, 2\pi).  \end{array}\right.
\end{eqnarray}

Equation (\ref{dLz}) is then reduced to
\begin{align}
\frac{d}{dt}L^{\sigma}_z(t)=U_{\sigma} \Big({\bar n}^{L}_{\sigma}(t)-\bar{n}^{R}_{\sigma}(t)\Big)\vartheta(\tau-t),
\label{Lz-time}
\end{align}
as shown in the main text.  In Eq.(\ref{Lz-time}), the definition of $\bar{n}^{L}_{\sigma}(t)$ and $\bar{n}^{R}_{\sigma}(t)$ can be found in Eqs.(\ref{nL}) and (\ref{nR}).  % denote the time-dependent average density of the Fermi gas with spin $\sigma$ when $\theta \in [0, 2\pi- \Delta \theta] $ and $\theta \in (2\pi-\Delta \theta, 2\pi)$, respectively.  %Here $n_{\sigma}(\theta,t)=\psi^{\dag}_{\sigma}(\theta,t) \psi_{\sigma}(\theta,t)$ is the density of a Fermi gas with spin $\sigma$.
Equation (\ref{Lz-time}) shows that $U_{\sigma}(\theta)$ breaks the conservation of $L^{\sigma}_z(t)$.  %which can be understood as follows.
As discussed in the main text, angular momentum is introduced into the system
through the density depletions induced by $U_{\sigma}(\theta)$, which leads to
$\bar{n}^{L}_{\sigma}(t)  >\bar{n}^{R}_{\sigma}(t)$.
For $t >\tau$,  when we turn off the potential, we have $ dL^{\sigma}_{z}(t)/dt=0$.

\subsection{Interacting case} \label{interaction-case}
Analogous to the noninteracting case,  when taking  an $s$-wave interaction between two spin species into consideration, the effective Hamiltonian under the mean-field approximation is given by $  H_{\rm MF}(t)= \int d\theta {\cal H}_{\rm eff}(\theta,t)$, with
\begin{align}
{\cal H}_{\rm eff}(\theta, t)=&\frac{|\Delta(\theta,t)|^2}{g}+\sum_{\sigma}\psi^{\dag}_{\sigma}(\theta,t)  {\cal H}_{\sigma}(\theta, t)\psi_{\sigma}(\theta,t) \nonumber\\
&+\left (\Delta(\theta,t)\psi^{\dag}_{\uparrow}(\theta,t)\psi^{\dag}_{\downarrow}(\theta,t)+{\rm h.c.}\right).
\label{Hmft}
\end{align}
Here $\psi_{\sigma}(\theta,t)$ satisfies
\begin{align}
i \hbar \frac{\partial }{\partial t} \left[\begin{array}{c}\psi_{\uparrow}(\theta,t) \\ \psi^{\dag}_{\downarrow}(\theta,t)\end{array}\right]
=\left[\begin{array}{cc} {\cal H}_{\uparrow}(\theta,t) & \Delta(\theta,t) \\  \Delta^{\ast}(\theta,t) & -{\cal H}_{\downarrow}(\theta,t)\end{array}\right]
\left[\begin{array}{c}\psi_{\uparrow}(\theta,t) \\ \psi^{\dag}_{\downarrow}(\theta,t)\end{array}\right].
\end{align}
Based on  $L^{\sigma}_z(t)=\langle \psi_{\sigma}(\theta,t) |{\hat L}_z |\psi_{\sigma}(\theta,t)\rangle$ and the time evolution of $\psi_{\sigma}(\theta,t)$,  the equation of motion for $L^{\sigma}_z(t)$ is modified to
\begin{align}
    \frac{d}{dt}L^{\uparrow}_z(t) &= U_{\uparrow} \Big({\bar n}^{L}_{\uparrow}(t)-\bar{n}^{R}_{\uparrow}(t)\Big)\vartheta(\tau-t)+\alpha(t),  \label{L1}\\
        \frac{d}{dt}L^{\downarrow}_z(t) &= U_{\downarrow} \Big({\bar n}^{L}_{\downarrow}(t)-\bar{n}^{R}_{\downarrow}(t)\Big)\vartheta(\tau-t)+\beta(t),  \label{L2}
 \end{align}
where we define
%\begin{widetext}
\begin{align}
\alpha(t) \equiv   \frac{1}{i \hbar }\int d\theta   &\left\{   \psi^{\dag}_{\uparrow}(\theta,t){\hat L}_z\Big[\Delta(\theta,t)\psi^{\dag}_{\downarrow}(\theta,t)\Big] \right. \nonumber\\
& \left.-\psi_{\downarrow}(\theta,t)\Delta^{\ast}(\theta,t){\hat L}_z \psi_{\uparrow}(\theta,t)\right\}, \\
\beta (t) \equiv  \frac{1}{i \hbar }\int d\theta &  \left\{  \psi_{\uparrow}(\theta,t) \Delta^{\ast}(\theta,t) \hat{L}_z \psi_{\downarrow}(\theta,t) \right. \nonumber\\
& \left. -\psi^{\dag}_{\downarrow}(\theta,t)\hat{L}_z \Big[\Delta(\theta,t)\psi^{\dag}_{\uparrow}(\theta,t)\Big]\right\}.
\label{alpha}
\end{align}
%\end{widetext}
In Eqs.(\ref{L1}) and (\ref{L2}), we observe that besides the single-particle term ${\cal H}_{\sigma}(\theta,t)$, the interaction also breaks the conservation of $L^{\sigma}_z(t)$ for $\alpha(t)\neq 0$ and $\beta(t)\neq 0$ in general. This is because interactions couple two spin components and introduce the exchange of angular momentum between the two species.

Although $\alpha(t) \neq 0$ and $\beta(t) \neq 0$ generally, we find that $\alpha(t)+\beta(t)=0$ is always satisfied. This can be demonstrated as follows.
Since $\Delta(\theta,t)=g \langle \psi_{\uparrow}(\theta,t) \psi_{\downarrow}(\theta,t)\rangle $, we have $\Delta^{\ast}(\theta,t)=g\langle \psi^{\dag}_{\downarrow}(\theta,t) \psi^{\dag}_{\uparrow}(\theta,t)\rangle$, so that $\alpha(t)+\beta(t)$ reduces to
\begin{align}
 \alpha(t)+\beta(t) =\frac{1}{g}|\Delta(\theta,t)|^2 \Big|^{\theta=2\pi}_{\theta=0}=0.
\label{alpha2}
\end{align}
In Eq. (\ref{alpha2}), we have considered $|\Delta(0,t)|=|\Delta(2\pi,t)|$. Equation (\ref{alpha2}) clearly shows that the interactions conserve the total angular momentum, although the conservation of angular momentum for a specific spin component is broken. The vanishing of $\alpha(t)+ \beta(t)$ satisfies our expectation, especially when $t>\tau$. This is because, after turning off $U_{\sigma}(\theta)$, the system becomes isolated, no exchange of angular momentum occurs between the system and environment, resulting in the conservation of the total angular momentum of the system.

\section{Renormalizing the bare interaction}
\label{renormal}
The renormalization relation of the bare interaction strength $g$ can be determined by solving a two-body problem. The full Hamiltonian of a Fermi gas  in a ring geometry is given by
\begin{align}
H=\sum_{m \sigma}\epsilon_m a^{\dag}_{m \sigma}a_{m\sigma}-\frac{g}{2\pi}\sum_{mm'k}a^{\dag}_{m+k, \uparrow}a^{\dag}_{m'-k,\downarrow}a_{m' \downarrow}a_{m\uparrow}
\end{align}
in the angular momentum space. Here, $\epsilon_m= m^2 \hbar^2/(2M R^2)$, and $a_{m \sigma} $ ($a^{\dag}_{m\sigma}$) denotes the  annihilation (creation) operator for a Fermi atom with spin $\sigma$ and angular momentum $m\hbar$.  The two-body bound state is $|\Psi \rangle=\sum_{m} \Phi_{m}a^{\dag}_{m \uparrow}a^{\dag}_{-m, \downarrow} |\rm vac \rangle$ and based on the Schr\"{o}dinger equation,  $H |\Psi \rangle =E_B | \Psi\rangle$, we have
\begin{align}
&\sum_{m} \left(2\epsilon_m-E_B\right)\Phi_{m}a^{\dag}_{m \uparrow}a^{\dag}_{-m, \downarrow} |{\rm vac } \rangle \nonumber\\
=&\frac{gC}{2\pi}\sum_{m}a^{\dag}_{m \uparrow}a^{\dag}_{-m, \downarrow}  |{\rm vac} \rangle,
\label{psim}
\end{align}
with $E_B$ $( E_B \leqslant 0)$ the binding energy of two-body bound state and  $C=\sum_{m} \Phi_{m}$.
From Eq. (\ref{psim}), we have
\begin{align}
\Phi_{m}=\frac{gC}{2\pi}\frac{1}{2\epsilon_m-E_B}.
\label{Phim}
\end{align}
Summing over $m$ in Eq. (\ref{Phim}), we have
\begin{align}
\frac{1}{g}=\frac{1}{2\pi}\sum_{m}\frac{1}{2\epsilon_m-E_B}.
\end{align}

\section{Bogoliubov-de Gennes formalism}
\label{TBdG}
\subsection{Dynamical  BdG formalism}
The dynamical BdG equations can be constructed as follows. We define the time-dependent field operator as
\begin{align}
\psi_{\sigma}(\theta,t)  =  \sum_{n}u_{\sigma n}(\theta,t)\gamma_{n \sigma}- s v^{\ast}_{\sigma n}(\theta,t)\gamma^{\dag}_{n \bar{\sigma}}, \label{psi-time}
\end{align}
where $\gamma_{n \sigma}$ and $\gamma^{\dag}_{n \sigma}$ are the annihilation and creation operators of static quasiparticle with energy $\epsilon_{n \sigma}$, which can be determined in the static BdG framework. Here, the time-dependent Bogoliubov coefficients $u_{\sigma n}(\theta,t)$ and $v_{\sigma n}(\theta,t)$ satisfy Eq. (\ref{td-BdG}), as shown in the main text,  according to the Heisenberg equation.
Considering the definition of the order parameter, we have
\begin{align}
\Delta(\theta,t)
= g\sum_{n}u_{\uparrow n}(\theta,t)v^{\ast}_{\downarrow n}(\theta,t)\vartheta(\epsilon_{n \uparrow}).
\label{Delta-time}
\end{align}
Thus,  when the initial states $u_{\sigma n}(\theta,t=0)$ and $v_{\sigma n}(\theta,t=0)$,  as well as the total particle number  $N$, are given,  the time evolution of $\Delta(\theta,t)$ can be self-consistently determined  from Eq. (\ref{td-BdG}) and Eq. (\ref{Delta-time}).

Solving Eq. (\ref{td-BdG}) generally requires a specific basis.  Here, we expand
$u_{ \uparrow n}(\theta, t) = \sum_m c_{nm}(t)\Theta_m(\theta)$ and  $v_{\downarrow n}(\theta,t )  =  \sum_{m}d_{nm}(t)\Theta_m(\theta)$
with $\Theta_m(\theta)=e^{i m \theta}/\sqrt{2\pi}$.  Thus the dynamical BdG equation in the $m$ space  reads
\begin{align}
i \hbar \frac{\partial }{\partial t} \left[\begin{array}{c}c_{nm}(t) \\ d_{nm}(t)\end{array}\right]
=\sum_{m^{\prime}} {\cal M}_{m,m'}(t) \left[\begin{array}{c}c_{nm^{\prime}}(t) \\ d_{nm^{\prime}}(t)\end{array}\right],
\label{time-BdG-matrix}
\end{align}
where
\begin{align}
{\cal M}_{m,m'}(t)&= \left[\begin{array}{cc}  {\cal H}^{m, m^{\prime}}_{\uparrow}(t) & \Delta_{m,m^{\prime}}(t) \\  \Delta^{\ast}_{m^{\prime},m}(t) & - {\cal H}^{m,m^{\prime}}_{\downarrow}(t)\end{array}\right],
\end{align}
with
\begin{align}
{\cal H}^{m,m^{\prime}}_{\sigma}(t)  &=  \left [\frac{m^2 \hbar^2}{2M R^2}-\mu_{\sigma}\right ]\delta_{mm^{\prime}}+f_{\sigma}(m,m^{\prime})\vartheta(t-\tau),    \\
\Delta_{m,m^{\prime}}(t)  &= \frac{1}{2\pi} \int d\theta \Delta(\theta,t)e^{i(m'-m)\theta}.
\end{align}
Here $f_{\sigma}(m,m') =\frac{1}{2\pi} \int  d\theta U_{\sigma}(\theta)e^{i(m'-m)\theta}$, which can be analytically expressed as
\begin{align}
f_{\sigma}(m,m') =\left\{\begin{array}{l} \frac{U_{\sigma}}{2} , \,\,\,\,\, m=m',\\
\frac{U_{\sigma}}{\Delta \theta (2\pi-\Delta \theta)}\frac{1-e^{i (m-m')\Delta \theta}}{(m-m')^2}, \,\,  m \neq m' .
\end{array}\right.
\label{fsigma-mmp}
\end{align}
Based on Eq.(\ref{time-BdG-matrix}) and Eq.(\ref{Delta-time}), when the initial states are given, $c_{nm}(t)$ and $d_{nm}(t)$ can be obtained. In this work,  we care about the time evolution of the angular momentum $L^{\sigma}_z(t)$,  which is given by
\begin{align}
L^{\uparrow}_{z}(t)  &=  \sum_{n,m}  (m\hbar)|c_{nm}(t)|^2 \vartheta(-\epsilon_{n \uparrow}), \\
L^{\downarrow}_{z}(t)  &=   \sum_{n,m}  (-m \hbar)|d_{nm}(t)|^2 \vartheta(\epsilon_{n \uparrow}).
\end{align}
%and the total angular momentum is $L^{\rm tot}_z(t)=\sum_{\sigma}L^{\sigma}_z(t)$.

\subsection{Static BdG formalism}
When turning to  the static BdG equation,  Eq.(\ref{td-BdG}) is reduced to
\begin{align}
\left[\begin{array}{cc}{\cal H}_{\uparrow}(\theta) & \Delta(\theta) \\ \Delta^{\ast}(\theta) & -{\cal H}^{\ast}_{\downarrow}(\theta)\end{array}\right]\left[\begin{array}{c}u_{\uparrow n}(\theta) \\ v_{\downarrow n}(\theta)\end{array}\right]=\epsilon_{n \uparrow}\left[\begin{array}{c}u_{\uparrow n}(\theta) \\ v_{\downarrow n}(\theta)\end{array}\right],
\label{HBdG}
\end{align}
and the self-consistent equations are
\begin{align}
\Delta(\theta)  &= g\sum_{n}u_{\uparrow n}(\theta)v^{\ast}_{\downarrow n}(\theta)\vartheta(\epsilon_{n \uparrow } ),  \label{self-eq1}\\
n_{\uparrow} (\theta)  &=  \sum_{n}|u_{\uparrow n}(\theta)|^2 \vartheta(-\epsilon_{n \uparrow }), \label{self-eq2}\\
n_{\downarrow}(\theta) &=  \sum_{n}|v_{\downarrow n}(\theta)|^2 \vartheta(\epsilon_{n \uparrow }).
\label{self-eq3}
\end{align}
Analogously to the dynamical BdG formalism,  we expand $u_{\uparrow n}(\theta)  =  \sum_m c_{nm}\Theta_m(\theta)$ and $
v_{\downarrow n}(\theta) =  \sum_{m}d_{nm}\Theta_m(\theta)$, and the static BdG equation  in the $m$ space becomes
\begin{align}
\sum_{m'}
{\cal M}_{m,m'}\left[\begin{array}{c}c_{nm'} \\d_{nm'}\end{array}\right]
=\epsilon_{n \uparrow}\left[\begin{array}{c}c_{nm} \\ d_{nm}\end{array}\right],
\label{cnm-dnm}
\end{align}
where
\begin{align}
{\cal M}_{m,m'}=\left[\begin{array}{cc}{\cal H}^{m,m'}_{\uparrow} & \Delta_{m,m'} \\ \Delta^{\ast}_{m',m}& -{\cal H}^{m,m'}_{\downarrow}\end{array}\right],
\end{align}
with
\begin{align}
 {\cal H}^{m,m'}_{\sigma}  &=  \left[ \frac{m^2 \hbar^2}{2MR^2}-\mu_{\sigma}\right] \delta_{mm'} ,  \\
 \Delta_{m,m'} &=  \frac{1}{2\pi} \int d\theta \Delta(\theta)e^{i(m'-m)\theta}.
\end{align}
Diagonalizing Eq.(\ref{cnm-dnm}), $c_{nm}$ and $d_{nm}$ can be obtained. Then from Eqs. (\ref{self-eq1}) to (\ref{self-eq3}),  $\Delta(\theta)$ and $n_\sigma(\theta)$ can be obtained self-consistently.

%%%%%%%%%%%%%%%%%%%%%%%%%%%%%%%%%%
\begin{figure}[t]
\begin{center}
\includegraphics[width=0.33\textwidth]{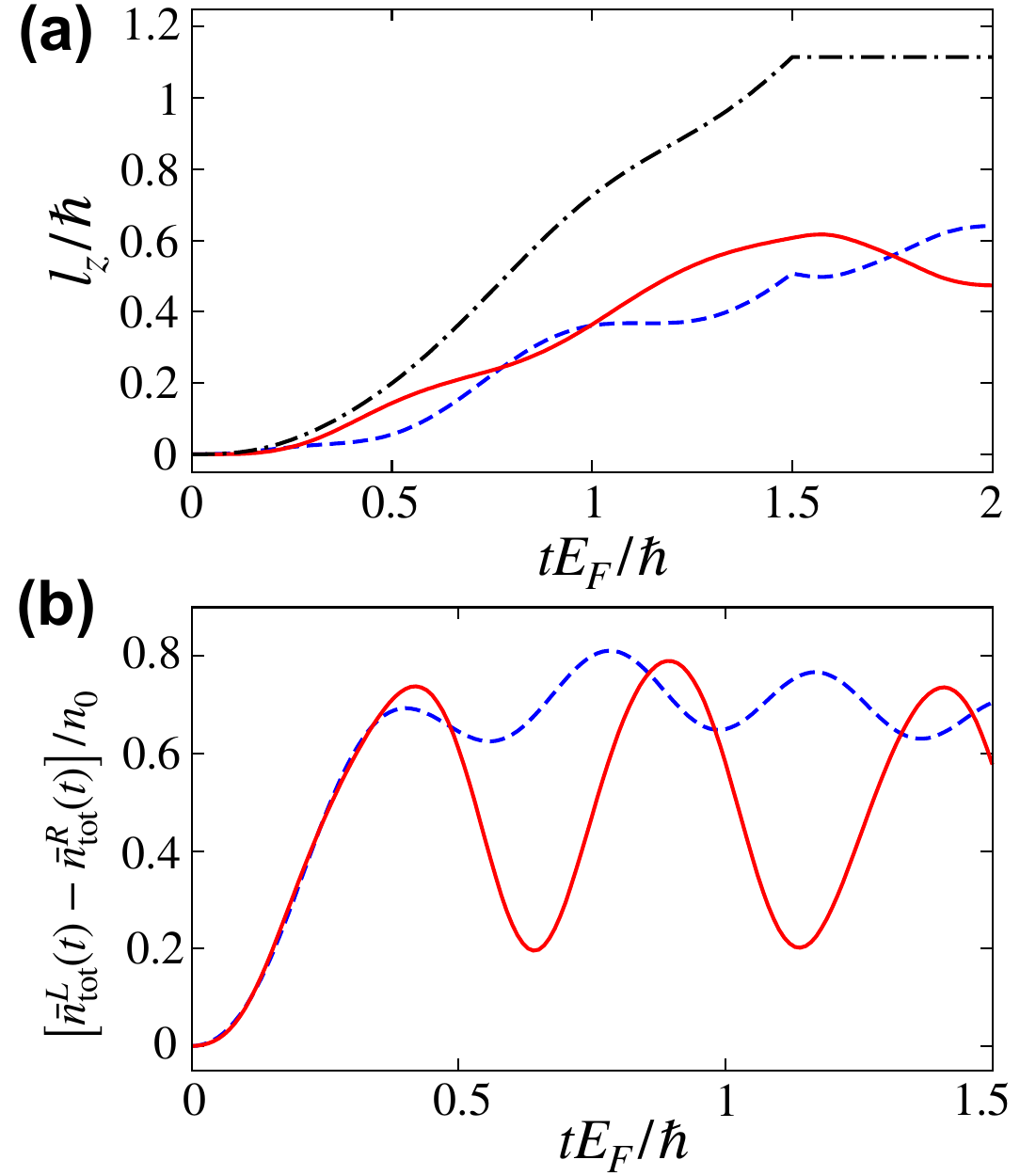}
\caption{(a) Evolution of $l^{\rm tot}_z(t)/\hbar$ (black dash-dotted curve), $l^{\uparrow}_z(t)/\hbar$ (blue dashed curve), and  $l^{\downarrow}_z(t)/ \hbar $ (red solid curve)  when $U_{\uparrow} \neq U_{\downarrow}$.  Here, $U_{\uparrow}/E_F=10$, $U_{\downarrow}/E_F=0$. (b) Evolution of  $[\bar{n}^{L}_{\rm tot}(t)-\bar{n}^{R}_{\rm tot}(t)]/n_0$ when $E_B/E_F=-5$ (blue dashed curve) and $E_B/E_F=-15$ (red solid curve), with $\bar{n}^{L(R)}_{\rm tot}(t)=\sum_{\sigma}\bar{n}^{L(R)}_{\sigma}(t)$.  Other parameters are the same as those in Fig.\ref{Fig2}.}
\label{FigS1}
\end{center}
\end{figure}
%%%%%%%%%%%%%%%%%%%%%%%%%%%%%%%%%%

%\section{Calculating the phase of $\Delta(\theta,t)$}

\section{Calculating the phase of $\Delta(\theta,t)$}
\label{phase of Delta}
The phase of $\Delta(\theta,t)$ can be extracted from the current $j_{\Delta}(\theta,t)$, defined as
\begin{align}
j_{\Delta}(\theta,t) & =  \Delta^{\ast}(\theta,t) \frac{\partial }{\partial \theta}\Delta(\theta,t)-\text{h.c.},
\label{j_Delta}
\end{align}
where $\Delta(\theta,t)$ can be  expressed as $\Delta(\theta,t)=|\Delta(\theta,t)|e^{i \phi(\theta,t)}$, as shown in the main text. Substituting this into Eq.(\ref{j_Delta}),  we obtain
$ \partial  \phi(\theta,t)/ \partial \theta  =  j_{\Delta}(\theta,t)/ (2i |\Delta(\theta,t)|^2)$.
Therefore, the phase difference $\Delta\phi(\theta,t)$ can be expressed as
\begin{align}
\Delta\phi(\theta,t) &=\phi(\theta,t)-\phi(0,t) =  \int^{\theta}_0 d\theta \frac{j_{\Delta}(\theta,t)}{2i |\Delta(\theta,t)|^2},
\end{align}
and $\Delta \phi(\theta,t)$ can be numerically calculated.

\section{Exchange of angular momentum for $U_{\uparrow} \neq U_{\downarrow}$}
\label{DiffU}

%\section{Exchange of angular momentum for $U_{\uparrow} \neq U_{\downarrow}$}

We consider $U_{\uparrow}\neq U_{\downarrow}$, and choose the BCS state as the initial state. Figure \ref{FigS1}(a) shows the evolutions of $l^{\sigma}_z(t)$ and $l^{\rm tot}_z(t)$. As illustrated in Fig.\ref{FigS1}(a), there exists angular momentum exchanges between the two spin components, which are different from the noninteracting case even though $U_{\uparrow}\neq U_{\downarrow}$.  The exchange of angular momentum is due to the fact that interactions couple the two spin components, thus transferring angular momentum between  them.

Here we also observe that when $t> \tau$, $l^{\rm tot}_z(t)$ becomes conserved, since interactions conserve the total angular momentum as demonstrated before. However, the exchange of angular momentum between the two spin components still exists. Such a behavior can be understood from the equations of motion for $L^{\sigma}_z(t)$ in the presence of interactions.

%\section{Effects of interactions}

\section{Effects of interactions}
\label{Diffint}

As discussed in the main text,  interactions favor homogenous density and suppress $l^{\rm tot}_z(t)$. Here we numerically confirm the above statement. As depicted in Fig.\ref{FigS1}(b), the stronger the interactions are, the smaller $\bar{n}^{L}_{\rm tot}(t)-\bar{n}^{R}_{\rm tot}(t)$ becomes. Based on Eq.(\ref{Lz-time}), we can find that for stronger interactions, the transfer of angular momentum is suppressed.

\bibliographystyle{apsrev4-2}

\begin{thebibliography}{99}
\bibitem{Tinkham-04}
M. Tinkham, \textit{Introduction to Superconductivity} (Dover Publications, Inc., Mineola, NY, 2004).

\bibitem{Annett-06}
 J. F. Annett, \textit{Superconductivity, Superfluids and Conden-
sates} (Oxford University Press Inc., New York, 2006).

\bibitem{Akkermans-07}
E. Akkermans and G. Montambaux, \textit{Mesoscopic Physics of Elec
trons and Photons} (Cambridge University Press, Cambridge,
England, 2007).

%Experiments of applications of superflow.
\bibitem{Tsai-99}
Y. Nakamura, Y. A. Pashkin, and J. S. Tsai,
Coherent control of macroscopic quantum states in a single-Cooper-pair box,
Nature (London) {\bf 398}, 786 (1999).

\bibitem{Wilhelm-08}
J. Clarke and F. K. Wilhelm,
Superconducting quantum bits,
Nature (London) {\bf 453}, 1031(2008).


%Significant experiment in SC.
\bibitem{Fairbank-61}
 B. S. Deaver and W. M. Fairbank, 
 Experimental Evidence for Quantized Flux in Superconducting Cylinders,
 Phys. Rev. Lett. {\bf 7}, 43 (1961).

 \bibitem{Doll-61}
 R. Doll and M. Näbauer, 
 Experimental Proof of Magnetic Flux Quantization in a Superconducting Ring,
 Phys. Rev. Lett. {\bf 7}, 51 (1961).

\bibitem{Yang-61}
N. Byers and C. N. Yang, 
Theoretical Considerations Concerning Quantized Magnetic Flux in Superconducting Cylinders,
Phys. Rev. Lett. {\bf 7}, 46 (1961).

\bibitem{Bloch-65}
 F. Bloch, 
 Off-Diagonal Long-Range Order and Persistent Currents in a Hollow Cylinder,
 Phys. Rev. {\bf 137}, A787 (1965).

 \bibitem{Onsager-61}
L. Onsager, 
Magnetic Flux Through a Superconducting Ring,
Phys. Rev. Lett. {\bf 7}, 50 (1961).


%synthetic gauge field.
\bibitem{Lin-09}
Y.-J. Lin, R. L. Compton, K. Jimenez-Garcia, J. V. Porto, and I. Spielman,
Synthetic magnetic fields for ultracold neutral atoms,
Nature (London) {\bf 462}, 628 (2009).

\bibitem{Dalibard-11}
J. Dalibard, F. Gerbier, G. Juzeliunas, and P. Öhberg,
Colloquium: Artificial gauge potentials for neutral atoms,
Rev. Mod. Phys. {\bf 83}, 1523 (2011).

\bibitem{Spielman-14}
N. Goldman, G. Juzeliūnas, P. Öhberg, and I. B. Spielman,
Light-induced gauge fields for ultracold atoms,
Rep. Prog. Phys. {\bf 77}, 126401 (2014).

%Rotate BEC and Fermi gase.
\bibitem{Cornell-99}
M. R. Matthews, B. P. Anderson, P. C. Haljan, D. S. Hall,
C. E. Wieman, and E. A. Cornell, 
Vortices in a Bose-Einstein Condensate,
Phys. Rev. Lett. {\bf 83}, 2498 (1999).

\bibitem{Madison-00}
K. W. Madison, F. Chevy, W. Wohlleben, and J. Dalibard,
Vortex Formation in a Stirred Bose-Einstein Condensate,
Phys. Rev. Lett. {\bf 84}, 806 (2000).

\bibitem{Ketterle-01}
J. Abo-Shaeer, C. Raman, J. Vogels, and W. Ketterle, 
Observation of Vortex Lattices in Bose-Einstein Condensates,
Science {\bf 292}, 476 (2001).

%Reviews of rotating Gases.

\bibitem{Fetter-09}
A. Fetter, 
Rotating trapped Bose-Einstein condensates,
Rev. Mod. Phys. {\bf 81}, 647 (2009).

\bibitem{Cooper-08}
N. Cooper, 
Rapidly rotating atomic gases,
Adv. Phys. 57, 539 (2008).

\bibitem{Ketterle-05}
M. W. Zwierlein, J. R Abo-Shaeer, A. Schirotzek, C. H. Schunck, and W. Ketterle,
Vortices and superfluidity in a strongly interacting Fermi gas,
Nature (London) {\bf 435}, 1047 (2005).

\bibitem{Piazza-09}
 F. Piazza, L. A. Collins, and A. Smerzi, 
 Vortex-induced phase-slip dissipation in a toroidal Bose-Einstein condensate flowing through a barrier,
 Phys. Rev. A {\bf 80}, 021601 (2009).

\bibitem{Phillips-13}
K. C. Wright, R. B. Blakestad, C. J. Lobb, W. D. Phillips, and G. K. Campbell,
Driving Phase Slips in a Superfluid Atom Circuit with a Rotating Weak Link,
Phys. Rev. Lett. {\bf 110}, 025302 (2013).


\bibitem{Cornell-14}
P. Engels, I. Coddington, P. C. Haljan, V. Schweikhard, and E. A. Cornell,
Observation of Long-Lived Vortex Aggregates in Rapidly Rotating Bose-Einstein Condensates,
Phys. Rev. Lett. {\bf 90}, 170405 (2014).

\bibitem{Wright-22}
Y. Cai, D. G. Allman, P. Sabharwal, and K. C. Wright,
Persistent Currents in Rings of Ultracold Fermionic Atoms,
Phys. Rev. Lett. {\bf 128}, 150401 (2022).


%Raman process. Using LG laser.
%LG laser.
\bibitem{Phillips-06}
M. F. Andersen, C. Ryu, P. Clad\'{e}, V. Natarajan, A. Vaziri, K. Helmerson, and W. D. Phillips,
Quantized Rotation of Atoms from Photons with Orbital Angular Momentum,
Phys. Rev. Lett. {\bf 97}, 170406 (2006).

\bibitem{Phillips-07}
C. Ryu, M. F. Andersen, P. Clade, V. Natarajan, K. Helmerson, and W. D. Phillips,
Observation of Persistent Flow of a Bose-Einstein Condensate in a Toroidal Trap,
Phys. Rev. Lett. {\bf 99}, 260401 (2007).

\bibitem{Phillips-11}
A. Ramanathan, K. C. Wright, S. R. Muniz, M. Zelan, W.T. Hill, C.J. Lobb, K. Helmerson, W.D. Phillips, and G. K. Campbell,
Superflow in a Toroidal Bose-Einstein Condensate: An Atom Circuit with a Tunable Weak Link,
Phys. Rev. Lett. {\bf 106}, 130401 (2011).

\bibitem{Hadzibabic-13}
S. Beattie, S. Moulder, R. J. Fletcher, and Z. Hadzibabic,
Persistent Currents in Spinor Condensates,
Phys. Rev. Lett. {\bf 110}, 025301 (2013).

\bibitem{Lin-18}
H.-R. Chen, K.-Y. Lin, P.-K. Chen, N.-C. Chiu, J.-B. Wang, C.-A. Chen, P.-P. Huang, S.-K. Yip, Y. Kawaguchi, and Y.-J. Lin,
Spin–Orbital-Angular-Momentum Coupled Bose-Einstein Condensates,
Phys. Rev. Lett. {\bf 121}, 113204 (2018).

\bibitem{Jiang-19}
D. Zhang, T. Gao, P. Zou, L. Kong, R. Li, X. Shen, X.-L. Chen, S.-G. Peng, M. Zhan, H. Pu, and K. Jiang,
Ground-State Phase Diagram of a Spin-Orbital-Angular-Momentum Coupled Bose-Einstein Condensate,
Phys. Rev. Lett. {\bf 122}, 110402 (2019).


%%%%%SOAMC theory.
%Bose gases.
\bibitem{Pu-15}
M. DeMarco and H. Pu,
Angular spin-orbit coupling in cold atoms,
Phys. Rev. A  {\bf 91}, 033630 (2015).

\bibitem{Sun-15}
K. Sun, C. Qu, and C. Zhang,
Spin–orbital-angular-momentum coupling in Bose-Einstein condensates,
Phys. Rev. A {\bf 91}, 063627 (2015).

\bibitem{Qu-15}
C. Qu, K. Sun, and C. Zhang,
Quantum phases of Bose-Einstein condensates with synthetic spin–orbital-angular-momentum coupling,
Phys. Rev. A {\bf 91}, 053630 (2015).

\bibitem{Chen-16}
 L. Chen, H. Pu, and Y. Zhang,
 Spin-orbit angular momentum coupling in a spin-1 Bose-Einstein condensate,
 Phys. Rev. A {\bf 93}, 013629 (2016).

\bibitem{Hu-19}
 X.-L. Chen, S.-G. Peng, P. Zou, X.-J. Liu, and H. Hu,
 Angular stripe phase in spin-orbital-angular-momentum coupled Bose condensates,
 Phys. Rev. Res. {\bf 2}, 033152 (2020).

\bibitem{Chen-19}
K.-J. Chen, F. Wu, J. Hu, and L. He,
Ground-state phase diagram and excitation spectrum of a Bose-Einstein condensate with spin-orbital-angular-momentum coupling,
Phys. Rev. A {\bf 102}, 013316 (2020).

\bibitem{Pu-20}
L. Chen, Y. Zhang, and H. Pu,
Spin-Nematic Vortex States in Cold Atoms,
Phys. Rev. Lett. {\bf 125}, 195303 (2020).

\bibitem{Duan-20}
Y. Duan, Y. M. Bidasyuk, and A. Surzhykov,
Symmetry breaking and phase transitions in Bose-Einstein condensates with spin–orbital-angular-momentum coupling,
Phys. Rev. A {\bf 102}, 063328 (2020).

%Fermi parts.
\bibitem{Chen-20}
K.-J. Chen, F. Wu, S.-G. Peng, W. Yi, and L. He, 
Generating Giant Vortex in a Fermi Superfluid via Spin-Orbital-Angular-Momentum Coupling,
Phys. Rev. Lett. {\bf 125}, 260407 (2020).

\bibitem{Wang-21}
 L.-L. Wang, A.-C. Ji, Q. Sun, and J. Li,
 Exotic Vortex States with Discrete Rotational Symmetry in Atomic Fermi Gases with Spin-Orbital–Angular-Momentum Coupling,
 Phys. Rev. Lett. {\bf 126}, 193401 (2021).



\bibitem{Chen-22}
K.-J. Chen, F. Wu, L. He, and W. Yi,
Angular topological superfluid and topological vortex in an ultracold Fermi gas,
Phys. Rev. Res.  {\bf 4}, 033023 (2022).

\bibitem{Han-22}
Y. Han, S.-G. Peng, K.-J. Chen, and W. Yi,
Molecular state in a spin–orbital-angular-momentum coupled Fermi gas,
Phys. Rev. A {\bf 106}, 043302 (2022).

\bibitem{Peng-22}
S.-G. Peng, K. Jiang, X.-L. Chen, K.-J. Chen, P. Zou, and L. He, 
Spin-orbital-angular-momentum-coupled quantum gases,
AAPPS Bulletin {\bf 32}, 36 (2022).

%Bose gases, Phase imprinting.
 \bibitem{Hadzibabic-12}
S. Moulder, S. Beattie, R.P. Smith, N. Tammuz, and Z. Hadzibabic,
Quantized supercurrent decay in an annular Bose-Einstein condensate,
Phys. Rev. A {\bf 86}, 013629 (2012).

 \bibitem{Kumar-18}
A. Kumar, R. Dubessy, T. Badr, C. De Rossi, M. de Go\"{e}r de Herve, L. Longchambon, and H. Perrin,
Producing superfluid circulation states using phase imprinting,
Phys. Rev. A  {\bf 97}, 043615 (2018).

%Fermi gases, Phase imprinting.
\bibitem{Roati-22}
G. Del Pace, K. Xhani, A. Muzi Falconi, M. Fedrizzi, N. Grani, D. Hernandez Rajkov, M. Inguscio, F. Scazza, W. J. Kwon, and G. Roati,  
Imprinting Persistent Currents in Tunable Fermionic Rings,
Phys. Rev. X
{\bf 12}, 041037 (2022).

\bibitem{Roati24}
L. Pezz\'{e}, K. Xhani, C. Daix, N. Grani, B. Donelli, F. Scazza, D. Hernandez-Rajkov, W. J. Kwon, G. Del Pace, and G. Roati,
Stabilizing persistent currents in an atomtronic Josephson junction necklace,
Nat. Commun. {\bf 15}, 4831 (2024).

\bibitem{Roati-23}
K. Xhani, G. Del Pace, F. Scazza, and G. Roati,
Decay of Persistent Currents in Annular Atomic Superfluids,
Atoms {\bf 11},109 (2023).


\bibitem{LO-65}
A. I. Larkin and Y. N. Ovchinnikov, 
Nonuniform state of superconductors,
Sov. Phys. JETP {\bf 20}, 762 (1965).

%\bibitem{supp}
%See Supplemental Material for details.

\bibitem{Xhani-24}
K. Xhani, A. Barresi, M. Tylutki, G. Wlazlowski, and P. Magierski,
Stability of persistent currents in superfluid fermionic rings,
arXiv:2406.10088 (2024).


\end{thebibliography}
\normalem
%imprinting1009.tex
%\bibliographystyle{apsrev4-1}

%Books of superconductor and superflow.

\end{document}